# QoS-Aware Power Minimization of Distributed Many-Core Servers using Transfer Q-Learning

Dainius Jenkus*, Fei Xia*, Rishad Shafik*, and Alex Yakovlev*
*School of Engineering, Newcastle University, Newcastle upon Tyne, NE2 1AD, UK
E-mail: {d.jenkus1, fei.xia, rishad.shafik, alex.yakovlev} @newcastle.ac.uk

*Abstract*—Web servers scaled across distributed systems necessitate complex runtime controls for providing quality of service (QoS) guarantees as well as minimizing the energy costs under dynamic workloads. This paper presents a QoS-aware runtime controller using horizontal scaling (node allocation) and vertical scaling (resource allocation within nodes) methods synergistically to provide adaptation to workloads while minimizing the power consumption under QoS constraint (i.e., response time). A horizontal scaling determines the number of active nodes based on workload demands and the required QoS according to a set of rules. Then, it is coupled with vertical scaling using transfer Q-learning, which further tunes power/performance based on workload profile using dynamic voltage/frequency scaling (DVFS). It transfers Q-values within minimally explored states reducing exploration requirements. In addition, the approach exploits a scalable architecture of the many-core server allowing to reuse available knowledge from fully or partially explored nodes. When combined, these methods allow to reduce the exploration time and QoS violations when compared to model-free Q-learning. The technique balances design-time and runtime costs to maximize the portability and operational optimality demonstrated through persistent power reductions with minimal QoS violations under different workload scenarios on heterogeneous multi-processing nodes of a server cluster.

## I. Introduction

An important aspect of web servers is the need to scale and support the user traffic demands at a specified quality of service (QoS) with minimal energy costs [1]. The compute node allocation and/or management of application instances is often provided by horizontal scaling using rule-based controls, which employ server-level metrics, e.g., CPU utilization [2], application workload characteristics [3] to make scaling decisions. Model-based techniques (e.g., using time-series analysis [2], neural networks [4]) have been proposed to improve the optimality of runtime controls. However, the portability is affected as extensive modelling data is required, which can be limited considering the complexity of decision space for distributed systems and dynamic environments of applications.

Alternatively, model-free Q-learning has been used [2], [5] for providing adaptability with little or no domain- or platform-specific knowledge. Although low-cost, the model-free horizontal scaling using Q-learning often suffer from much greater QoS violation rates during the exploration. At higher complexity, reinforcement learning (RL) hybrid controls [6] and model-assisted Q-learning [7] have been deployed for achieving faster learning and lower QoS violations. Authors of [8] proposed energy minimization using transfer learning approach, where learnt DVFS actions are mapped to unexplored states achieving faster learning and better application performance. This work balances design-time and runtime costs to maximize the portability and operational optimality by making the following ***contributions***:

- a QoS-aware runtime controller (termed as RHQV Scaler) with joint horizontal and vertical scaling to provide elastic controls that minimize power cost and QoS violations under dynamic workloads,
- vertical scaling using the principle of transfer Q-learning combining proposed Inter-Node Learning Transfer (INLT) and Intra-State Learning Transfer (ISLT) methods to minimize QoS violations,
- experimental validation and comparative analysis of the method, demonstrating minimal QoS violations and minimized power under different workload scenarios.

The remainder of this paper is organized as follows: Section II explains the proposed runtime management. A case study and discussions are presented in Section III.

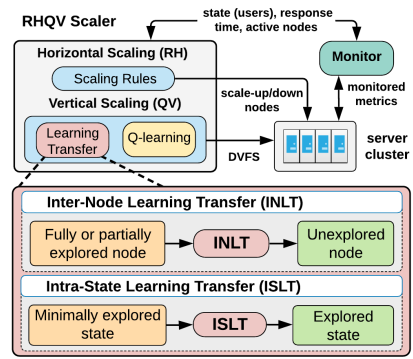

Fig. 1: Runtime controller (RHQV Scaler) and its interfaces.

## II. Proposed Runtime Management

The runtime controls are carried out by rule-based horizontal scaling (RH) and vertical scaling using transfer Q-learning (QV) together termed as RHQV Scaler (Fig. 1). A monitor collects metrics, e.g., users, response time (*rt*) to guide scaling decisions and allow to carry out learning at runtime.

***Rule-based Horizontal Scaling:*** firstly, a hierarchical controls of RHQV synergistically tunes power/performance of an application by scaling up/down nodes using a set of rules, while ensuring QoS (defined as a soft response time constraint). The scale-up action is reactively triggered and enables an additional node when QoS is violated and vertical scaling is no longer sufficient to maintain the required QoS. As workload decreases, the scale-down rule ensures that *rt* is within the QoS constraint and that the workload would be sustained after scaling-down to avoid instability of scaling actions. This is achieved by comparing a current state (defined as active users) to the available history of states when the system made a transition to a current configuration of nodes.

*Vertical Scaling using Transfer Q-learning*: the power/performance of an application is further tuned by globally scaling CPU frequencies of active nodes using DVFS (see Fig. 1). The controls are orchestrated by the Q-learning agent, where the Q-values are estimated as follows:

$$Q(s,a) = Q(s,a) + \alpha[R(s,a) + \gamma maxQ'(s',a') - Q(s,a)], \quad (1)$$

where $R(s,a)$ is the reward, $\alpha$ is the learning rate, and $\gamma$ is the discount factor applied for the next state $s'$. The Q-values are transferred using the learning transfer consisting of ISLT and INLT components.

At the early exploration stage of an unexplored application, ISLT identifies how the performance varies (e.g., linearly) with different DVFS scaling actions across multiple states (i.e., active users). Once identified, the exploration of minimally explored states is guided to select specific actions, e.g., min/max DVFS operating points in order to derive the state's performance-action relationship as an approximated function, $f_{rt}$. The Q-values are then transferred using eq. 1, where the reward is estimated using eq. 2. For each unexplored action within a state, the reward is found using a predicted response time from the identified performance-action function (i.e., $f_{rt}$). The best rewards are given to actions allowing the system to operate neither too fast (wasting energy), nor to slow (violating the QoS). The reward is defined as follows:

$$R(s,a) = \begin{cases} \beta_0 \times (T_{slack}/T_{rt}), & \text{if } T_{slack} < 0, \\ \beta_1 \times (1/T_{slack}), & \text{otherwise}, \end{cases} \quad (2)$$

where $T_{rt}$ is the QoS constraint, $T_{slack} = T_{rt} - rt$, $\beta_1$ is the constant for scaling the positive rewards ($T_{slack} < 0$), while $\beta_0$ scales the negative rewards received when QoS is violated.

INLT works by exploiting scalability provided by the clusters consisting of identical compute nodes. It transfers available knowledge between nodes, which host instances of distributed applications. The INLT works by mapping an unobserved state in an unexplored multi-node configuration to a range of states, which correspond to a fully or partially explored reference node. Then, such state is searched in the Q-table of the reference node, and if found, the corresponding Q-values are transferred to unexplored nodes. Otherwise, the exploration continues as described in ISLT method.

### III. CASE STUDY AND DISCUSSIONS

The proposed controller is evaluated in terms of power savings and QoS violation rate (% of control intervals with violated QoS). The case study application is a Wordpress website hosted on a cluster of four Odroid XU4s. Two workload cases are investigated covering irregular web traffic (rapidly changing) and a scaled one-day English Wikipedia trace, which is applied using the developed workload generator. Fig. 2 (a) indicates that ISLT knowledge transfer has allowed to improve the selection of vertical scaling decisions (DVFS actions) reducing the QoS violations by around 50%, while exposed to both workloads for the first time. Fig. 2 (b) displays QoS violation rate when a second unexplored node is enabled and Q-values are transferred from explored node using INLT. Similarly to ISLT, it shows that QoS violations are lower for Wikipedia workload and are minimized approximately by a factor of two when compared to the model-free Q-learning.

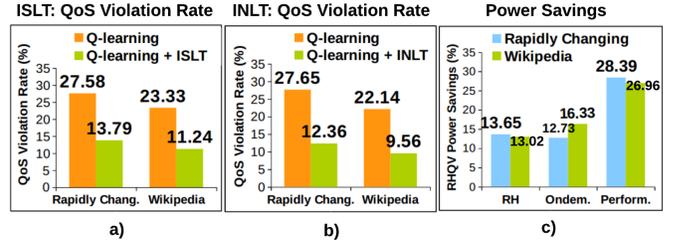

Fig. 2: Comparative QoS trade-offs and power savings.

The effectiveness of power minimization has been evaluated using QoS-unaware Linux *Ondemand* and *Performance* governors [9]. Also, *RH* method, which excludes vertical scaling but employs identical horizontal scaling to RHQV. The proposed scaler provides up to 28.39% savings when compared to the system using *Performance* governor and up to 16.33% over *Ondemand* governor as shown in Fig. 2 (c). RHQV offers 13.65% more savings when compared to *RH* (without vertical scaling) suggesting that the rest of power reductions come from the transfer Q-learning based vertical scaling. This highlights the advantage of having synergistically joint vertical and horizontal scaling controls for power minimization.